\documentclass[twocolumn,showpacs,preprintnumbers,amsmath,amssymb,superscriptaddress]{revtex4}

\usepackage{graphicx}
\usepackage{graphicx,epsf} 
\usepackage{dcolumn}
\usepackage{bm}
\usepackage{latexsym}
\newcommand{\xsize}{\epsfxsize=7.0cm}

\begin{document}

\title{Optimization of Robustness of Complex Networks}

\author{G. Paul,$^1$ T. Tanizawa,$^{1,2}$ S. Havlin,$^{1,3}$
  and H. E. Stanley}

\affiliation{Center for Polymer Studies and Dept. of Physics, Boston
             University, Boston, MA 02215, USA\\
$^2$Department of Electrical Engineering, Kochi National
             College of Technology\\ Monobe-Otsu 200-1, Nankoku, Kochi,
             783-8508, JAPAN\\
$^3$Minerva Center and Department of Physics, Bar Ilan University\\
Ramat Gan 52900,  Israel}
 
\begin{abstract}
Networks with a given degree distribution may be very resilient to one
type of failure or attack but not to another. The goal of this work is
to determine network design guidelines which maximize the robustness of
networks to both random failure and intentional attack while keeping the
cost of the network (which we take to be the average number of links per
node) constant.  We find optimal parameters for: (i) scale free networks
having degree distributions with a single power-law regime, (ii)
networks having degree distributions with two power-law regimes, and
(iii) networks described by degree distributions containing two peaks.
Of these various kinds of distributions we find that the optimal network
design is one in which all but one of the nodes have the same degree,
$k_1$ (close to the average number of links per node), and one node is
of very large degree, $k_2 \sim N^{2/3}$, where $N$ is the number of
nodes in the network.

\end{abstract}


\maketitle

\section{Introduction}
Recently, there has been much interest in the resilience of real-world
networks to failure of nodes or to intentional attacks
\cite{Albert,Paxon,Cohen2000,Callaway,Cohen2001,Cohen2002}. The goal of
this work is to determine network design guidelines which maximize the
robustness of the networks to both random failures of nodes and attacks
targeted on the highest degree nodes \cite{notex}.

\begin{figure}[tbh]
\centerline{
\xsize
\epsfclipon
\epsfbox{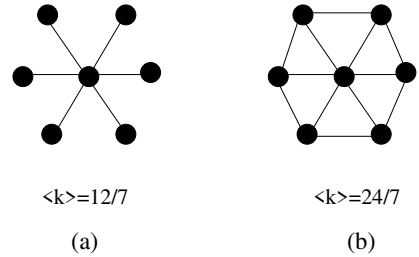}
}
\caption{(a) Example of network with low tolerance to targeted
  attack.  (b) Example of a network with much higher tolerance
  to targeted attack but with double the cost.}
\label{ex}
\end{figure}

Networks with a given degree distribution may be very resilient to one
type of attack but not to another. Consider the simple seven node
network example shown in Fig.~\ref{ex}a. This network is relatively
robust with respect to a random failure - only a failure of the central
node will cause the network to fragment. Thus the probability that a
random failure will cause the network to fragment is only 1/7. On the
other hand the network is extremely vulnerable to a targeted attack - an
attack in which the most highly connected nodes are removed first. In
this simple example the probability that a targeted attack which removes
one node will fragment the network is 1!

As shown in Fig.~\ref{ex}b we can modify the network to make it more
resilient to targeted attack by adding more links between the nodes on
the periphery of the network. With this modification, neither a single
node random failure nor a targeted attack which removes only one node
can fragment the network. This increased robustness, however, comes with
a cost. If we define the ``cost'' to construct and maintain a network
with a given number of nodes as being proportional to the average number
of links $\langle k \rangle$ per node in the network, we see that the
cost of the original network is 12/7 while the cost of modified network
is 24/7. So for the additional robustness we pay a factor of 2 in cost.

Our goal then becomes how to maximize the robustness of a network of
size N nodes to both random failures and targeted attacks with the
constraint that the cost remains constant. That is, the number of links
remains constant but the nodes are connected in a different and more
optimal way.

Many real world computer, social, biological and other types of networks
have been found to be scale free, i.e., they exhibit degree
distributions of the form $P(k) \sim k^{-\lambda}$
~\cite{Bar99,Faloutsos,Barabasi,Broder,Ebel,Redner,Jeong,Mendes,RPT}.
For large scale free networks with exponent $\lambda$ less than 3, it has
been found that, if nodes fail randomly, essentially all nodes must fail
for the network to become disconnected~\cite{Cohen2000,Callaway}.  On the
other hand, because the scale free distribution has a long power-law
tail (i.e. hubs with large degree), the networks are vulnerable with
respect to targeted attack.  This raises two questions that we address
in this work: (i) How can we optimize scale free networks to both random
failure and targeted attack and (ii) Are there other network types that
can be better optimized than scale free networks.  To this end, we
begin our analysis with scale free networks and then consider networks
with other types of distributions.

\section{Optimization Metric} The threshold for random removal of nodes for
any degree distribution, $P(k)$, is \cite{Cohen2000}
\begin{equation}
\label{e1}
f_c^{\mbox{\scriptsize rand}}=1-{1\over\kappa_0-1},
\end{equation}
where $\kappa_0\equiv\langle k^2 \rangle/\langle k \rangle$.

Reference \cite{Cohen2001} describes how to calculate $f_c^{\mbox{\scriptsize targ}}$, the threshold under
intentional attack.

A metric we can use to measure the robustness of the network to both
random and targeted attack is the sum
\begin{equation}
\label{e9}
f_c^{\mbox{\scriptsize tot}}=f_c^{\mbox{\scriptsize rand}}
    +f_c^{\mbox{\scriptsize targ}}.
\end{equation}
This is only one of a number of possible metrics we could use, e.g., we
could have used the product $f_c^{\mbox{\scriptsize rand}}\cdot
f_c^{\mbox{\scriptsize targ}}$. Our results are, in
general, not dependent on the metric chosen.

Our goal can now be stated as follows: for a network of a given number
of nodes $N$, how do we maximize $f_c^{\mbox{\scriptsize tot}}$ while
keeping the number of links constant?

We can estimate an upper bound for $f_c^{\mbox{\scriptsize tot}}$. We
first note that the maximum value of $f_c^{\mbox{\scriptsize rand}}$ is
essentially 1 which is the case when a small number of nodes have a very
large degree distribution -- as in scale free networks with $\lambda<3$
or in the simplest case where one node is linked to all other nodes.  In
these cases the probability of these critical nodes randomly failing
approaches zero and the threshold is close to 1.  The maximum value of
$f_c^{\mbox{\scriptsize targ}}$ is obtained in the situation in which
all the nodes have the same degree, $\langle k \rangle$, in which case
the targeted attack becomes equivalent to random failure and we can use
Eq.~(\ref{e1}) to find $f_c^{\mbox{\scriptsize targ}}=1-1/(\langle k
\rangle - 1)$.  Our upper bound $\bar f_c^{\mbox{\scriptsize tot}}$ is
therefore given by
\begin{equation} 
f_c^{\mbox{\scriptsize tot}}\leq \bar f_c^{\mbox{\scriptsize tot}}\equiv 2-{1\over{(\langle k \rangle -1)}}.
\label{upperbound}
\end{equation}

\section{Power Law Degree Distribution} We first study how to optimize a scale free network with a single power
law regime by varying the exponent $\lambda$ and keeping $\langle k
\rangle$ constant~\cite{note0}.

\begin{figure}[tbh]      
\centerline{
\xsize
\epsfbox{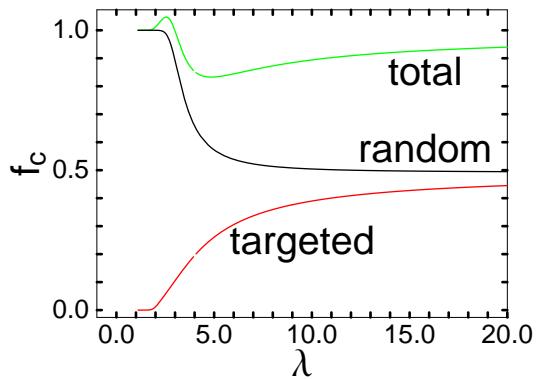}
}
\caption{Random, targeted and total critical percolation thresholds
  for scale free networks as a function of the  exponent $\lambda$.}
\label{pss}
\end{figure}

In Fig.~\ref{pss}, we plot the values of $f_c^{\mbox{\scriptsize
rand}}$, $f_c^{\mbox{\scriptsize targ}}$, and $f_c^{\mbox{\scriptsize
tot}}$, for a range of the exponent $\lambda$ for a network with
$N=10^6$ nodes and $\langle k \rangle =3$~\cite{note1}.  For this choice
of $\langle k \rangle$, the upper bound of $f_c^{\mbox{\scriptsize
tot}}$ is given by $\bar f_c^{\mbox{\scriptsize tot}} \approx 1.5$ [see
Eq.(~\ref{upperbound})].  We find that: As $\lambda$ increases,
$f_c^{\mbox{\scriptsize targ}}$ increases but $f_c^{\mbox{\scriptsize
rand}}$ decreases.  For $\lambda\approx 2.5$ $f_c^{\mbox{\scriptsize
tot}}$ is optimized but the maximum value of $f_c^{\mbox{\scriptsize
tot}}$ ($\approx 1.04$) is small relative to the theoretical maximum
$\approx 1.5$.  It is interesting that the network is optimized with a
value of $\lambda$ about 2.5 which is consistent with the range of
exponents for many real
networks~\cite{Bar99,Faloutsos,Barabasi,Broder,Ebel,Redner,Jeong,Mendes,RPT}.

\section{Degree Distributions Formed by Two Power Laws}

We next analyze a slightly more complex form for $P(k)$. Keeping
$\langle k\rangle$ constant, we consider degree distributions which
consist of 2 segments each of which is a power law. The inflection point
at which the distribution changes slope we denote by $a$.  The
hypothesis is that the first power law segment (for $k < a)$ with
exponent $\alpha$ will contribute to the robustness against targeted
attack and the second segment (for $k>a$) with exponent $\lambda$ will
contribute to the robustness against random failures.  We determine the
relative weights of the two segments such that $f_{\mbox{\scriptsize
c-total}}$ is maximized.  To maintain constant $\langle k \rangle$ as we
change $a$ we again adjust the minimum, $m$, of the distribution.

\begin{figure}[tbh]          
\centerline{ \xsize \epsfbox{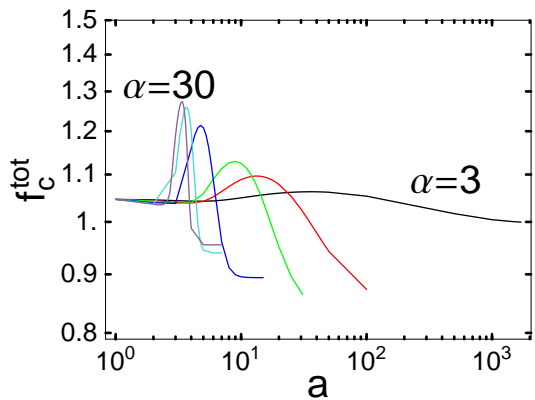} }
\caption{Total percolation threshold vs the inflection point $a$ for
  distributions $P(k)$ composed of two scale free segments with
  $\lambda=2.5$ (for $ k >a$) and for the slope (for $k < a$)
  $\alpha=3,4,5,10,20,30$ (from right to left).}
\label{c25}
\end{figure}

In Fig.~ \ref{c25}, we plot the values of $f_c^{\mbox{\scriptsize tot}}$
as a function of the inflection point $a$ for $\lambda=2.5$, $\langle
k\rangle=3$ and for various $\alpha$.  We see that
$f_c^{\mbox{\scriptsize tot}}$ attains a maximum value that increases
with increasing $\alpha$.  Thus for a given $\lambda$ we can maximize
$f_c^{\mbox{\scriptsize tot}}$ by choosing appropriate values of $a$ and
$\alpha$.

We can further increase the maximum value of $f_c^{\mbox{\scriptsize
tot}}$ by changing the value of $\lambda$. In plots (not shown) of
$f_c^{\mbox{\scriptsize tot}}$ as functions of $a$, for $\alpha=10$ and
various values of $\lambda$, we find that as $\lambda$ decreases, the
maximum value attained by $f_c^{\mbox{\scriptsize tot}}$ increases.
Thus we can maximize the robustness of a network with respect to both
random failure and targeted attack by replacing the original degree
distribution by one with the same $\langle k\rangle$ but with two power
law segments characterized by exponents $\alpha$ and $\lambda$ with
$\alpha$ large and $\lambda$ close to one (the lowest value of $\lambda$
which yields physical results).

In these distributions with large values of $\alpha$ the total
probability in the tail of the distribution is a small fraction of the
total probability, so that there is only on the order of one node in the
tail and, due to the large value of $\alpha$, most of the nodes have
almost the same number of links -- very close to the minimum $m$.

\section{Degree Distributions Formed by an Exponential and a Power Law} With
 the insight that the larger the exponent $\alpha$ the better the
 optimization, we now consider distributions with the initial power law
 segment ($k < a$) of the distribution replaced by an exponential
 distribution $P(k)\sim\epsilon^{-\beta k}$. As expected we find that
 for a given $\beta$, at some value of $a$, $f_c^{\mbox{\scriptsize
 tot}}$ is optimized and that the optimization increases as $\beta$
 increases.

\section{Degree Distribution Formed by Two Gaussians} Considering the
previous cases, it appears that the optimization strategy does not
depend on the fact that the initial segment of the distribution is a
power law or exponential.  Given that the total probability of the nodes
in the second segment (the tail of the distribution) is very small (of
order 1), as discussed above, we now want to study the case where the
second segment is not a  power law.  We therefore consider here a
case where the degree distribution consists of two Gaussian segments.

One Gaussian has its center at $k_1$ and width $\omega_1$ and the second
Gaussian has its center at $k_2 > k_1$ and width $\omega_2$.  The ratio
$r$ represents the fraction of the number of nodes in the second
Gaussian to the total number of nodes.  We consider cases in which $r$
and $k_2$ are the independent variables and $k_1$ must be a dependent
variable in order to maintain a fixed value of $\langle k \rangle$.

In plots (not shown) of the total threshold $f_c^{\mbox{\scriptsize tot}}$ in
terms of the ratio $r$ for various values of $k_2$, we find that the
optimal $f_c^{\mbox{\scriptsize tot}}$ increases and the optimal value
of $r$ decreases as the value of $k_2$ increases.  In addition, we
obtain higher values of the optimal $f_c^{\mbox{\scriptsize tot}}$ for
smaller values of $\omega_1$.  This fact indicates that the highest
value of $f_c^{\mbox{\scriptsize tot}}$ is achieved in the limit where
this width goes to zero.  In this limit the lower segment tends toward a
simple delta function.  This observation motivates us to study next the
optimization of networks consisting of two delta functions .

\section{Degree Distributions Formed by Two Delta Functions} Next we
consider the degree distribution that consists of two delta
functions:
\begin{equation}
P(k)\equiv (1-r) \delta (k - k_1) + r \delta (k - k_2).
\end{equation}
As in the case of two Gaussian segments, we calculate the total
threshold as a function of $r$ and $k_2$ for a fixed value of $\langle k
\rangle$.  We obtain analytical expressions for both
$f_c^{\mbox{\scriptsize rand}}$ and $f_c^{\mbox{\scriptsize targ}}$ as
follows.

Using Eq. ~(\ref{e1}),

\begin{equation}
f_c^{\mbox{\scriptsize rand}} = \frac{\langle k \rangle^2 - 2 r \langle k \rangle
k_2 - 2(1-r)\langle k \rangle + r k_2^2} {\langle k \rangle^2 - 2 r
\langle k \rangle k_2 - (1-r)\langle k \rangle + r k_2^2}.
\label{eq:fc-random-td}
\end{equation}

For the threshold for targeted attack, we must consider two cases:

(i)$f_c^{\mbox{\scriptsize targ}} > r$. In this case, after the targeted
attack, the only nodes that remain have degree $k_1$.  We find
\begin{equation}
f_c^{\mbox{\scriptsize targ}} = r + \frac{1 - r}{\langle k \rangle - r
k_2} \left\{ \langle k \rangle \frac{\langle k \rangle - r k_2 - 2
\left(1-r \right)} {\langle k \rangle - r k_2 - \left(1 -r \right)} - r
k_2 \right\}.
\label{eq:fc-targeted td upper}
\end{equation}

(ii) $f_c^{\mbox{\scriptsize targ}} < r$.  For this case nodes are
removed only from the higher segment and we find
\begin{equation}
f_c^{\mbox{\scriptsize targ}} = \frac{\langle k \rangle^2 - 2 r \langle k
\rangle k_2 + r k_2^2 - 2 (1-r) \langle k \rangle} {k_2 (k_2 - 1) (1 -
r)}.
\label{eq:fc-targeted td lower}
\end{equation}
With the expressions for the thresholds, Eqs.\ (\ref{eq:fc-random-td}),
(\ref{eq:fc-targeted td upper}), and (\ref{eq:fc-targeted td lower}), we
are able to evaluate the total threshold $f_c^{\mbox{\scriptsize tot}}$.
We can obtain an expression for the optimal value of $k_2$ as a function
of $r$ by determining the value of $k_2$ for which
$f_c^{\mbox{\scriptsize tot}}$ is maximized.  Based on our results
above, we expect the maximum will be obtained for $r$ small.  Using
Eqs.~(\ref{eq:fc-random-td}) and (\ref{eq:fc-targeted td lower}), we
find that for small $r$ the optimal value of $k_2$ can be approximated
by
\begin{equation}
k_2 \sim \left\{{2 \langle k \rangle^2  (\langle k \rangle -1)^2 \over
  {2\langle k\rangle-1}}\right\}^{1/3} r^{-2/3}\equiv A r^{-2/3}.
\label{k2r}
\end{equation}
Using this result and  Eq.~(\ref{upperbound}) we find, for small r,
\begin{equation}
f_c^{\mbox{\scriptsize tot}}=\bar f_c^{\mbox{\scriptsize tot}}-{3 \langle k \rangle \over A^2}
r^{1/3} + O(r^{2/3}).
\label{ftr}
\end{equation}

Thus $f_c^{\mbox{\scriptsize tot}}$ approaches the theoretical maximum
value when $r$ approaches, but is not, zero.  For a network of N nodes,
the maximum value of $f_c^{\mbox{\scriptsize tot}}$ is obtained when
$r$=$1/N$ the smallest possible value consistent with there being 1 node
of degree $k_2$.  Given this $r$ the equation determining the optimal
$k_2$ is
\begin{equation}
k_2=A N^{2/3}.
\label{k2}
\end{equation}

\begin{figure}[tbh]                       
\centerline{
\xsize
\epsfbox{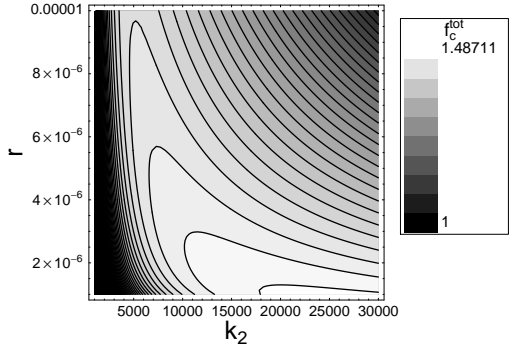}
}
\caption{Contour plot of total percolation threshold vs $r$ and $k_2$ for
distribution consisting of two delta functions with $\langle k
\rangle=3$.}
\label{contourTDkAv3}
\end{figure}

Figure~\ref{contourTDkAv3} demonstrates the behavior of the optimal
$f_c^{\mbox{\scriptsize tot}}$ as a function of $r$ and $k_2$.  We see
that the highest values of the optimal $f_c^{\mbox{\scriptsize tot}}$
are attained as $r$ approaches zero; and for a given small value of $r$,
$f_c^{\mbox{\scriptsize tot}}$ is optimized for $k_2$ from Eq.~(\ref{k2}).

The general nature of our results hold for the metric defined in
Eq.~(\ref{e9}) as well as for metrics $f_c^{\mbox{\scriptsize tot}}$
defined as a linear combination of the random and targeted thresholds
\begin{equation}
\label{metric1}
f_c^{\mbox{\scriptsize tot}}=a f_c^{\mbox{\scriptsize rand}}
+b f_c^{\mbox{\scriptsize targ}},
\end{equation}
where $a$ and $b$ allow one to specify for a given network the
importance to be attached to random and targeted attack respectively.
The only modification to our results for these alternative
metrics, is that the prefactor A is generalized to
\begin{equation}
A= \left\{ {a \over b} {2 \langle k \rangle^2 (\langle k \rangle -1)^2
  \over {2\langle k\rangle-1}}\right\}^{1/3}.
\label{metric2}
\end{equation}

\section{Discussion and Summary} 
\begin{figure}[tbh]                       
\centerline{
\xsize
\epsfbox{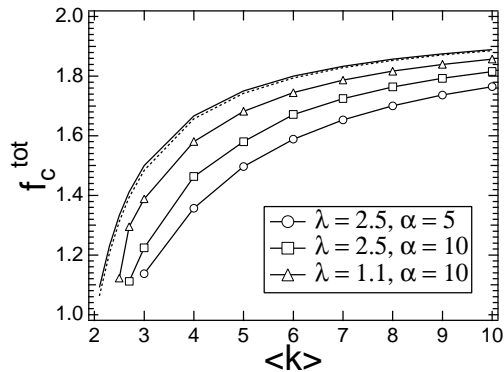}
}
\caption{Plots of the optimal  $f_c^{\mbox{\scriptsize tot}}$ vs $\langle k
  \rangle$ for theoretical maximum value (solid line), two delta
  functions (dotted line), and distributions consisting of two power
  law segments (see legend) 
 }
\label{comparison}
\end{figure}

We develop a strategy for optimization of
scale free and two-peaked networks against both random failures and
targeted attacks.  To our knowledge, this is the first study of the
robustness of complex networks to multiple types of failure/attack.  We
find that the network which approaches the theoretical maximum level of
optimization is generated with a degree distribution which is non-zero
at only two values: $k_1$ and $k_2$.  This level of optimization is
possible because in order to obtain a value of $f_c^{\mbox{\scriptsize
rand}}$ which is essentially 1 we have to wire only 1 node with a large
number of links. The remaining nodes, all with the same degree, provide
essentially the same high degree of resilience to targeted attack as for
the case in which all nodes have degree $\langle k \rangle$.
Figure~\ref{comparison} compares the level of optimization obtained for
these optimized networks two-delta-function networks with the level of
optimization obtained for networks with two power laws segments and with
the theoretical maximum values which can be obtained.

The optimal network is obtained by connecting $k_2 \sim A N^{2/3}$ nodes
to a single node and all of the other nodes except the degree $k_2$ node
are of degree $k_1 \sim \langle k \rangle - A/N^{1/3} \sim \langle k
\rangle$.

Subjects for further study include (i) an analysis of
the static and dynamic properties of the optimized two delta function
networks which we have identified here and (ii) the optimization of
complex networks under combined random failure and targeted
attack. Finally we note that the origin of the $N^{2/3}$ appearing in
Eq.~(\ref{k2}) may be related to the size of the infinite cluster at
criticality for Erd\"os-R\'enyi graphs \cite{Erdos59,Erdos60,Bollobas}.

We thank L. Braunstein, S. Buldyrev,and S. Sreenivasan for helpful
discussions and ONR and Israel Science Foundation for support.


\begin{thebibliography}{99}
\bibitem{note0} In order to keep $\langle k \rangle$ constant, as we
change $\lambda$ we also must change $m$, the minimal number of links a
node can have.  Note that the value of $m$ decreases as we increase
$\lambda$.
\bibitem{note1} We obtain similar results for other values of $\langle k \rangle$.
\end{thebibliography}
\end{document}